\documentclass[aps,twocolumn,prb]{revtex4}
\usepackage{amssymb}
\usepackage{graphicx}
\begin{document}
\title{LONG RANGE COHERENT MANIPULATION OF NUCLEAR SPINS IN QUANTUM HALL FERROMAGNET}
\author{T.Maniv$^{1,2}$ , Yu. A. Bychkov$^{1,3}$, I.D. Vagner$^{1,4}$}
\date{\today{}}

\affiliation{$^{1}$Grenoble High Magnetic Field Laboratory,Max-Planck-\\
Institute fur Festkorperforschung and CNRS, Grenoble\\
France.\\
$^{2}$Chemistry Department, Technion-Israel Institute of\\
Technology, Haifa 32000, Israel\\
$^{3}$L.D.Landau Institute for Theoretical Physics, Kosygina 2, Moscow,Russia\\
$^{4}$Department of Communication Engineering,\\
Holon Institute of Technology, Holon 58102, Israel}


\begin{abstract}
A coherent superposition of many nuclear spin states can be prepared and
manipulated via the hyperfine interaction with the electronic spins by
varying the Landau level filling factor through the gate voltage in
appropriately designed Quantum Hall Ferromagnet. During the manipulation
periods the 2D electron system forms spatially large Skyrmionic spin
textures, where many nuclear spins follow locally the electron spin
polarization. It is shown that the collective spin rotation of a single spin
texture is gapless in the limit of zero Zeeman splitting, and may dominate
the nuclear spins relaxation and decoherence processes in the quantum well.

\end{abstract}

\maketitle


\section{Introduction}

The emerging fields of quantum information processing and quantum computing
(QC) \cite{Steane98} have stimulated recently a flurry of activity in the
established fields of atomic and condensed matter physics, approaching
fundamental questions, such as the influence of measurement on quantum
mechanical systems or the meaning of phase coherence in interacting many
particle systems, from a stinkingly new point of view. \ Experimental
realization of QC has been so far successfully achieved, however, only in
devices consisting of a few qubits.

The idea presented in this paper should not be considered as a proposal for
building any kind of quantum computer. Instead it addresses the general
problem of how to store and manipulate a large number of qubits without
losing their phase coherence. \ This is done with respect to a concrete
physical system consisting of nuclear spins in semiconducting
heterojunctions under the conditions of the odd integer Quantum Hall (QH)
effect \cite{PVK98}. Our proposal has been motivated by the set of
experiments, reported in \cite{Barret95,Tycko95}, where the Knight shift, $%
K_{S}$, and the spin lattice relaxation time $T_{1}$ of the $^{71}$Ga nuclei
in GaAs multiple quantum well (MQW) structure under perpendicular magnetic
field were detected by means of the optically pumped NMR technique. The
electronic Landau level filling factor was varied in these experiments by
tilting the magnetic field axis with respect to the 2D layers. The Knight
shift was found to reduce dramatically as the filling factor was shifted
slightly away from $\nu =1$, indicating that the injection of a single
charge into the 2D electron system is followed by reversal of many
electronic spins. In the same interval of the filling factor the relaxation
time was found to drop by several orders of magnitude with respect to its
value in the quantum Hall ferromagnetic ground state.

Both effects are considered as strong evidence for the creation of
skyrmionic spin texture \cite{Girvin99} in the electronic spin distribution
as the filling factor shifts slightly away from unity, and indicate the
crucial importance of the hyperfine interaction in controlling the nuclear
spin dynamics. Since the hyperfine interaction is the dominant coupling of
the nuclear spins to their environment they may be exploited as quantum bits
(qubits), provided the environment, that is the 2D electron gas, is in a
nondissipative, coherent quantum state ( e.g. like the QH ferromagnetic
state at LL filling factor $\nu =1$ at low temperature \cite{MBVW01}).
Furthermore, as will be shown below, near $\nu =1$ it may be possible to
manipulate coherently a large number of nuclear spins through the hyperfine
interaction with the electronic spin texture by \textit{varying a single
parameter} , such as the Landau level filling factor, through changes in the
gate voltage.

At filling factor $\nu =1$ the ground state of the 2D EG is ferromagnetic
even in the limit of zero Zeeman energy \cite{Sondhi93}. Flipping nuclear
spins in this state through the hyperfine interaction is followed by the
creation of spin excitons \cite{BIE81},\cite{KH84}. The energy cost of this
excitation can be minimized if both the electron and the hole are created at
the nuclear position, where the energy gain associated with the e-h Coulomb
attraction is exactly compensated by the exchange energy of the hole. Yet,
the remaining small Zeeman energy (on the electronic energy scale) is a huge
energy gap for the nuclear spins. The extremely long spin-lattice relaxation
time observed by Barrett \textit{et al} \cite{Barret95,Tycko95} may be due
to this energy gap (see below, however). Overcoming the Coulomb attraction
by increasing the e-hole distance leads to increasing the exciton transverse
momentum. The corresponding excitation energy scales with the Coulomb
energy, which is $\sim 100K$ , that is much larger than the Zeeman
splitting. The spin exciton spectrum is strongly influenced by long range
electrostatic potential fluctuations, which can trap the electron and the
hole separately in local potential wells and so reduce, or even completely
remove the energy gap \cite{BMVW94}.

Slightly away from $\nu =1$ the lowest energy state of the electron gas is a
spin texture, in which the average spin distribution is smoothly twisted in
space in order to minimize the exchange energy \cite{Sondhi93}. The size of
the twist is determined by the Zeeman energy \cite{Fertig94,BMV96,BKMV98}.
Microscopic calculations, based on Hartree-Fock (HF) approximation for a
single, isolated Skyrmion \cite{Fertig94,MFB96,FBCMKS97}, have found a
family of low energy excitations, with an approximately quadratic relation
between the energy and the number of flipped spins,$K$ , which can be
associated with the kinetic rotational energy of the entire spin texture
about its symmetry axis. However, except for the special case where $K$ is a
half integer, the spectrum has an excitation gap, which is some fraction of
the large Coulomb energy scale. To account for the observed enhancement of
the nuclear relaxation rate, these authors have suggested \cite{CMBFGS97}
that at filling factor slightly away from $\nu =1$, where there is a finite
density of Skyrmions, the ground state is a Skyrme crystal, for which the
spin waves spectrum is gapless due to the breakdown of the global spin
rotation symmetry. This appealing interpretation is hard to reconcile with
the latest optically pumped NMR (OPNMR) measurements \cite{Barrett01}. Based
on this OPNMR data, the many-Skyrmion state does not appear consistent with
the closed packed periodic lattice described in \cite{CMBFGS97}. Instead, it
was suggested \cite{Barrett01} that the Skyrmions' tail is drastically
reduced, e.g. due to the effect of disorder potential \cite{Nederveen99},
leading to some kind of spatially inhomogeneous state of nearly independent
pinned Skyrmions.\

This conclusion motivates us to carefully reexamine the problem of spin
excitations in a single Skyrmion\cite{Fertig94,MFB96,FBCMKS97}. Our study
has shown that the excitation gap in the collective rotational spectrum of a
single Skyrmion goes to zero when the Skyrmion radius tends to infinity, and
that for the characteristic Skyrmion sizes found experimentally the gap is a
small fraction of the Zeeman energy scale, rather than of the large Coulomb
energy scale, as claimed previously. \ This is done within the framework of
a phenomenological approach, similar to that taken by Girvin \textit{et al }
\cite{Girvin99}, which is based on microscopic HF calculation. The influence
of these low-lying electron spin excitations on the nuclear spin
polarization and phase coherence via the hyperfine interactions is then
discussed.

\section{The model}

We start our analysis by considering the Hamiltonian for nuclear spins
interacting with 2D electron gas in MQW structure
\begin{equation}
\widehat{H}=-\hbar \gamma _{n}\sum_{j}\widehat{\mathbf{I}}_{j}\cdot \mathbf{B%
}_{0}-\hbar \gamma _{e}\int d^{2}r\widehat{\mathbf{S}}\left( \mathbf{r}%
\right) \cdot \mathbf{B}_{0}+\widehat{H}_{ee}+\widehat{H}_{en}
\label{Hamiltonian}
\end{equation}
where
\begin{equation}
\widehat{H}_{en}=A\sum_{j}\widehat{\mathbf{S}}\left( \mathbf{r}_{j}\right)
\cdot \widehat{\mathbf{I}}_{j}  \label{Hint}
\end{equation}
Here $\widehat{\mathbf{I}}_{j}$ is the nuclear spin operator located at $%
\mathbf{r}_{j}$ , $\widehat{\mathbf{S}}\left( \mathbf{r}\right) $ is the
electronic spin density operator, $\mathbf{B}_{0}$ is the external magnetic
field, which is assumed to be oriented perpendicular to the 2D electron gas
( $\mathbf{B}_{0}=B_{0}\mathbf{z}$ ) , $\widehat{H}_{ee}$ is the
electron-electron interaction, $\gamma _{n}=g_{n}\mu _{n}/\hbar $ $\ $and $%
\gamma _{e}=g_{e}\mu _{B}/\hbar $ the nuclear and electronic gyromagnetic
ratios respectively, and $A=\frac{8\pi }{3}g_{n}\mu _{n}g_{0}\mu _{B}\left|
u_{0}\left( 0\right) \right| ^{2}$ is the Fermi contact hyperfine coupling
constant. In this expression $u_{0}\left( 0\right) $ is the periodic part of
the Bloch wavefunction at the nucleus, and $g_{0}$ is the g-factor of a free
electron. We use the standard normalization \ $\int_{\upsilon }\left|
u_{0}\left( \mathbf{r}\right) \right| ^{2}d^{3}r=\upsilon $, where $\upsilon
$ is the volume of a unit cell in the crystal.

The manipulation of the nuclear spins is carried out through spin flip-flop
processes, associated with the 'transverse' part of the interaction
Hamiltonian $\widehat{H}_{en}$ ( Eq.(\ref{Hint}) ) , i.e. $\frac{1}{2}%
A\sum_{j}\left[ \widehat{I}_{j,+}\widehat{S}_{-}\left( \mathbf{r}_{j}\right)
+\widehat{I}_{j,-}\widehat{S}_{+}\left( \mathbf{r}_{j}\right) \right] $ ,
where $\widehat{I}_{j,+}=\widehat{I}_{j,x}+i\widehat{I}_{j,y},\widehat{I}%
_{j,-}=\widehat{I}_{j,x}-i\widehat{I}_{j,y}$ , and $\widehat{S}_{+}\left(
\mathbf{r}\right) =\widehat{\psi }_{\uparrow }^{\dagger }\left( \mathbf{r}%
\right) \widehat{\psi }_{\downarrow }\left( \mathbf{r}\right) $ , $\widehat{S%
}_{-}\left( \mathbf{r}\right) =\widehat{\psi }_{\downarrow }^{\dagger
}\left( \mathbf{r}\right) \widehat{\psi }_{\uparrow }\left( \mathbf{r}%
\right) $. Here $\widehat{\psi }_{\sigma }\left( \mathbf{r}\right) ,\widehat{%
\psi }_{\sigma }^{\dagger }\left( \mathbf{r}\right) $ are the electron field
operators with spin projections $\sigma =\uparrow ,\downarrow $.

The strength of the hyperfine coupling constant can be estimated by using
the expression

\begin{equation}
K_{S}\equiv \frac{1}{h}A\langle \widehat{S}_{z}\left(
\mathbf{r}_{j}\right) \rangle \approx \alpha \left( n_{2D}/2\pi
l\right)  \label{KS}
\end{equation}
for the Knight shift at filling factor $\nu =1$, where $\alpha \equiv
A/\hbar $ , $n_{2D}$ is the areal density of the 2D electron gas, and $l$ is
the width of the QW. For the $^{31}$Ga nucleus (with $g_{n}\approx .27$) in
GaAs $\left| u_{0}\left( 0\right) \right| ^{2}\sim 10^{4}$, and for the
parameters characterizing the sample used by Barrett \textit{et al} \cite
{Barret95},i.e. $l\approx 30\text{nm}$ , and $n_{2D}=1.5\times 10^{11}\text{%
cm}^{-2}$ , one finds $K_{S}\sim 10^{4}\text{Hz}$ , in good agreement with
Ref. \cite{Barret95}.

In the framework of the model just described, we will now show how, by
varying the LL filling factor, a large number of nuclear spins can be
prepared in a state appropriate to start quantum computation. A number, $n$
, stored in the memory of a hypothetical quantum computer made of nuclear
spins, may be described as a direct product of $N$ pure nuclear spin states
\[
\left| n\right\rangle =\left| n_{1}\right\rangle \otimes \left|
n_{2}\right\rangle \otimes ...\otimes \left| n_{N}\right\rangle
\]
where $\left| n_{j}\right\rangle =\sum_{\sigma =\pm 1}\delta _{n_{j},\sigma
}\left| j,\sigma \right\rangle $ , $\delta _{n_{j},\sigma }$ is the
Kronecker delta , and $\left| j,\sigma \right\rangle $ is a nuclear state
with spin projection $\sigma $ for a nucleus located at $\mathbf{r}_{j}$. To
carry out a quantum computing process, however, a coherent superposition of
such products , \ i.e. $\left| \psi \right\rangle =\sum_{n=1}^{N}\alpha
_{n}\left| n\right\rangle $, should be prepared at time $t=0$. This
superposition may be represented more transparently for our purposes by the
direct product of $N$ mixed spin up and spin down states,
\[
\left| \psi \left( t=0\right) \right\rangle =\prod_{j=1}^{N}\otimes \left(
u_{j}\left| j,\uparrow \right\rangle +v_{j}\left| j,\downarrow \right\rangle
\right)
\]
with the normalization $\left| u_{j}\right| ^{2}+\left| v_{j}\right| ^{2}=1$.

While the hyperfine coupling with the electron spins is the dominant
interaction of the nuclear spin qubits system with its environment, it is
only a weak perturbation to the electron spins system. Thus, at a
temperature which is much lower than any electronic energy scale in this
system, the electronic spins at LL\ filling factor $\nu $ should be in the
corresponding ground state,$\left| 0;\nu \right\rangle $. One may,
therefore, construct an effective nuclear spin Hamiltonian by projecting the
combined nuclear-electronic spin Hamiltonian, Eq.(\ref{Hamiltonian}), on the
ground electronic state, $\left| 0;\nu \right\rangle $. The resulting
effective nuclear spin Hamiltonian can be written as:
\[
\widehat{\mathcal{H}}_{n}=-\hbar \gamma _{n}\sum_{j=1}^{N}\widehat{\mathbf{I}%
}_{j}\cdot \mathbf{B}_{0}+A\sum_{j=1}^{N}\mathbf{S}\left( \mathbf{r}%
_{j}\right) \cdot \widehat{\mathbf{I}}_{j}
\]
where $\mathbf{S}\left( \mathbf{r}\right) =\left\langle 0;\nu \right|
\widehat{\mathbf{S}}\left( \mathbf{r}\right) \left| 0;\nu \right\rangle $ is
the expectation value of the electronic spin density in the ground
electronic state at filling factor $\nu $. \

The corresponding state of the nuclear spin system can be found by
considering $u_{j}$ and $v_{j}$ as variational parameters, and then
minimizing the energy functional $\mathcal{E}_{n}\mathcal{=}\left\langle
\psi \right| \widehat{\mathcal{H}}_{n}\left| \psi \right\rangle $,with
respect to $u_{j}$ , $v_{j}$ . As noted above , at $\nu =\nu _{0}\neq 1$ , $%
\mathbf{S}\left( \mathbf{r}\right) $ has nonzero transverse components,
associated with the skyrmionic spin texture, smoothly varying in space.

A simple calculation shows that
\[
\mathcal{E}_{n}\mathcal{=}\frac{1}{2}\sum_{j}\left\{ \Omega _{j}\left(
\left| v_{j}\right| ^{2}-\left| u_{j}\right| ^{2}\right) +\left[
Av_{j}u_{j}^{\ast }S_{-}\left( \mathbf{r}_{j}\right) +c.c\right] \right\}
\]
\ where $\Omega _{j}=\gamma _{n}B_{0}-\alpha S_{z}\left(
\mathbf{r}_{j}\right) $ is the local nuclear Zeeman energy. The
extremum conditions ( subject to the normalization $\left|
u_{j}\right| ^{2}+\left| v_{j}\right| ^{2}=1$ ) are readily solved
to yield: $\left| v_{j}\right|
^{2},\left| u_{j}\right| ^{2}=\frac{1}{2}\left( 1\pm \frac{\hbar\Omega _{j}}{%
\varepsilon _{j}}\right) $ ,\ \ where $\varepsilon _{j}=\sqrt{\hbar
^{2}\Omega _{j}^{2}+A^{2}\left| S_{+}\left( \mathbf{r}_{j}\right) \right|
^{2}}$ . In this state the nuclear spin polarization $\left\langle \psi
\right| \widehat{\mathbf{I}}_{j}\left| \psi \right\rangle $ follows the
underlying electronic spin texture; the transverse component takes the form
\begin{equation}
I_{j,+}=\left\langle \psi \right| \widehat{I}_{j,+}\left| \psi \right\rangle
=u_{j}^{\ast }v_{j}=\pm \frac{1}{2}AS_{+}\left( \mathbf{r}_{j}\right)
/\varepsilon _{j}  \label{I+(r)}
\end{equation}
whereas the longitudinal component is
\[
I_{j,z}=\left\langle \psi \right| \widehat{I}_{j,z}\left| \psi
\right\rangle =\frac{1}{2}\left( \left| u_{j}\right| ^{2}-\left|
v_{j}\right| ^{2}\right) =\pm \frac{1}{2}\hbar\Omega
_{j}/\varepsilon _{j}
\]

Thus, the nuclear spins distribution follows the distribution of the
electronic Skyrmion spin texture. The key parameter here is the local mixing
parameter
\[
\eta _{j}\equiv \left( A/\hbar \Omega _{j}\right) \left| S_{+}\left( \mathbf{%
r}_{j}\right) \right| =\left( 2\pi K_{S}/\Omega _{j}\right) \left|
\widetilde{S}_{+}\left( \mathbf{r}_{j}\right) \right|
\]
with $\widetilde{S}_{+}\left( \mathbf{r}_{j}\right) \equiv \frac{S_{+}\left(
\mathbf{r}_{j}\right) }{\left( n_{2D}/l\right) }$ ,\ \ which determines the
local deviation of the nuclear spins state from a pure ferromagnet. Thus,
for $\eta _{j}\ll 1$ the many nuclear spin state is very close to a pure
ferromagnet. In the opposite extreme limit, $\eta _{j}\gg 1$ , all
individual nuclear spin states are equally probable, i.e. $\left|
v_{j}\right| ^{2}=\left| u_{j}\right| ^{2}\rightarrow 1/2$ , and so one
generates an ideal starting state for quantum computing \cite{Unruh95}. As
we shall see below, this extremely strong mixing condition is not realistic.
In the intermediate situation, where $\eta _{j}\sim 1$ almost everywhere,
the distributions $\left| v_{j}\right| ^{2},\left| u_{j}\right| ^{2}$ vary
moderately around the mean value $1/2$.

The condition for achieving such a desired situation is, therefore, two
folded : (1) The average Knight shift, $K_{S}$ , should be comparable to the
average nuclear Zeeman frequency, $\Omega $ , i.e. $\left( 2\pi K_{S}/\Omega
\right) \sim 1$; and (2) the transverse component of the normalized
electronic spin density, $\left| \widetilde{S}_{+}\left( \mathbf{r}%
_{j}\right) \right| $, should be of the order one over a large spatial
region (namely a region consisting of many nuclear spins ). Usually the
Knight shift is a small fraction of the NMR frequency, so that the first
condition is not easily fulfilled. An exceptional example will be discussed
toward the end of the paper. The second condition is satisfied by large
skyrmionic spin texture (i.e. for sufficiently small effective g-factor).

Let us now outline very briefly a scenario for manipulating many
nuclear spins in MQW by varying the LL filling factor. Very fast
changes of the filling factor can be achieved without overheating
the nuclear spins system, by varying the gate voltage. The process
might start at an early time, $t=-\tau _{0}$, when the filling
factor was tuned at $\nu =\nu _{0}$ , slightly away from $\nu =1$,
and then kept fixed until $t=0$. If the 'waiting' time $\tau _{0}$
is much longer than the ( relatively short) relaxation time
$T_{1}\left( \nu =\nu _{0}\right) $, then at $t=0$ the nuclear
spins would be settled in their ground state corresponding to the
electron gas at filling factor $\nu =\nu _{0}$. \ By so doing the
nuclear spin qubits are prepared in a state which is an
appropriate initial state for quantum computing. However, to
shield the nuclear spins from decoherence
due to the\ low-lying electronic spin fluctuations, which are present at $%
\nu =\nu _{0}\neq 1$, the filling factor may be quickly switched back to $%
\nu =1$ (i.e. on a time scale shorter than $T_{1}\left( \nu =\nu _{0}\right)
$), so that the nuclear spins are trapped in their mixed, textured state,
unable to relax for a long time to the pure ferromagnetic ground state
dictated by the electron gas at $\nu =1$, since $T_{1}\left( \nu =1\right) $
is extremely long. \ \

\section{Collective mode}

\ As discussed above, during the manipulation cycle, when the nuclear spins
have relatively short relaxation and dephasing times, their dynamics is
controlled by the low-lying spin fluctuations of the electron gas through
the hyperfine interaction. For a single, isolated Skyrmion the rigid
rotation of the entire spin texture about its symmetry ($Z$) axis is a zero
mode, which can be responsible for such low energy fluctuations. The
generator of this rotation, $\widehat{L}_{z}$ , is the $Z$-component of the
angular momentum of the entire spin texture. To find a 'classical'
Hamiltonian for this rotational motion in the electronic spin space, one may
exploit the Hartree-Fock approximation for the Skyrmion energy near filling
factor $\nu =1$ , consisting of Coulomb +Zeeman +nonuniformity energy, that
is \cite{BKMV98}:
\begin{equation}
E_{tot}\left( R\right) =\frac{3\pi ^{2}e^{2}}{2^{6}\kappa R}+\frac{e^{2}}{%
4\kappa l_{H}}\sqrt{\frac{\pi }{2}}\left( \frac{R}{l_{sk}}\right) ^{2}\ln
\left( \frac{2l_{sk}}{R}\right)  \label{Etot}
\end{equation}
where $R$ is a variational parameter describing the Skyrmion core radius,\ $%
l_{sk}$\ is the length scale corresponding to the Skyrmion's tail, $%
l_{sk}^{-2}=2\sqrt{\frac{2}{\pi }}\left| g\right| \widetilde{a}%
_{B}/l_{H}^{3} $ , $\widetilde{a}_{B}=\kappa \hbar ^{2}/m_{0}e^{2}$ is the
effective Bohr radius ($m_{0}$ being the free electron mass, and $\kappa $-
the dielectric constant), and $l_{H}=\sqrt{c\hbar /eH}$-the magnetic length.

The Zeeman energy associated with the reversed spins is
\[
\Delta E_{Z}=g\mu _{B}H\widetilde{L}_{z}=\frac{e^{2}}{4\kappa l_{H}}\sqrt{%
\frac{\pi }{2}}\left( \frac{R}{l_{sk}}\right) ^{2}\ln \left( \frac{2l_{sk}}{%
\sqrt{\bar{e}}R}\right)
\]
where, $\widetilde{L}_{z}\equiv L_{z}/\hbar $ , and $\bar{e}$ stands for the
natural logarithm base, so that the total number of reversed electronic
spins in the Skyrmion is related to the core radius $R$ through the
expression:
\begin{equation}
\widetilde{L}_{z}=\left( \frac{R}{l_{H}}\right) ^{2}\ln \left( \frac{2l_{sk}%
}{\sqrt{\bar{e}}R}\right)  \label{SzVsR}
\end{equation}

Minimization with respect to $R$ yields for the equilibrium core radius:
\[
\frac{3\pi ^{2}e^{2}}{2^{6}\kappa R_{eq}^{3}}=\left( \frac{2}{l_{H}^{2}}%
\right) \ln \left( \frac{2l_{sk}}{R_{eq}}\right) \left| g\right| \mu _{B}H
\]
while the second derivative $\left[ \frac{\partial ^{2}}{\partial R^{2}}%
E_{tot}\right] _{eq}\approx \left( \frac{6\left| g\right| \mu _{B}H}{%
l_{H}^{2}}\right) \ln \left( \frac{2l_{sk}}{R}\right) $, or by Eq.(\ref
{SzVsR}):

\begin{equation}
U=\left[ \frac{\partial ^{2}}{\partial \widetilde{L}_{z}^{2}}E_{tot}\right]
_{eq}\approx \left( \left| g\right| \mu _{B}H\right) \left( \frac{l_{H}}{%
R_{eq}}\right) ^{2}\frac{3\ln \left( \frac{2l_{sk}}{R}\right) }{2\ln
^{2}\left( \frac{2l_{sk}}{\sqrt{\bar{e}}R}\right) }  \label{Ueq}
\end{equation}

Expanding the energy, Eq.(\ref{Etot}), up to second order in $\widetilde{L}%
_{z}$ about its equilibrium value, $K$, that is writing
\[
E_{tot}\left( \widetilde{L}_{z}\right) =E_{tot}\left( K\right) +\frac{1}{2}%
U\left( \widetilde{L}_{z}-K\right) ^{2}+...
\]
\ the second term on the RHS corresponds to the 'classical'
rotational energy of the entire spin texture about its symmetry
axis. At the classical level any deviation of $\widetilde{L}_{z}$
from its equilibrium value $K$ corresponds to a continuous
deformation (or more precisely a uniform contraction or expansion
) of the Skyrmion with respect to its equilibrium configuration,
thus conserving its topological charge, but increasing the
Skyrmion energy with respect to its equilibrium value. The
collective rotation of the Skyrmion in spin space is therefore
reflected as a radial expansion or contraction in orbital space.
Quantization of this rotational motion can be achieved by
replacing $L_{z}\rightarrow \frac{\hbar }{i}\frac{\partial
}{\partial \varphi }$ , where $\varphi $ is the rotation angle,
which yields \
\begin{equation}
\widehat{H}_{rot}=\frac{1}{2}U\left( \frac{1}{i}\frac{\partial }{\partial
\varphi }-K\right) ^{2}  \label{Hrot}
\end{equation}

Note that since $K=\left\langle 0;\nu \right| \widehat{L}_{z}\left| 0;\nu
\right\rangle /\hbar $ , its value usually does not coincide with any
(discrete) eigenvalue of the operator $\frac{1}{i}\frac{\partial }{\partial
\varphi }$, so that the spectrum of $\widehat{H}_{rot}$ has usually a gap of
the order of the rotational energy constant,$\ U$ . Remarkably the above
estimate, Eq.(\ref{Ueq}), shows that for a large Skyrmion, $R_{eq}\gg l_{H}$
, $U$ is a small fraction of the Zeeman energy $\varepsilon _{sp}=\left|
g\right| \mu _{B}H$ , that is $U\sim \varepsilon _{sp}\left( \frac{l_{H}}{%
R_{eq}}\right) ^{2}\ll \varepsilon _{sp}$. $\ $The fraction, $\left( \frac{%
l_{H}}{R_{eq}}\right) ^{2}$, tends to zero as the Skyrmion core radius
becomes macroscopic, reflecting the macroscopic inertial mass associated
with the collective rotation of a macroscopic spin texture. \ Using the
equilibrium value $R_{eq}$ as a function of the g-factor obtained above, we
find that \ $\left( \frac{l_{H}}{R_{eq}}\right) ^{2}\sim 2\left| \widetilde{g%
}\right| ^{2/3}$, where $\widetilde{g}\equiv g\left( \frac{\widetilde{a}_{B}%
}{l_{H}}\right) $.\ For a typical experimental value of the effective
electronic g-factor,\ $\widetilde{g}\sim .002$ , it is found that $U\sim
\allowbreak 3\times 10^{-2}\varepsilon _{sp}$\ .

It is interesting to note that the magnetic field dependence of $U$,
expressed by Eq.(\ref{Ueq}), indicates similarity of the collective
rotational motion to precession of a magnetic moment in a magnetic field.
Indeed, by equating the classical expression for the rotational energy, $%
H_{rot}=\frac{\hbar ^{2}}{2U}\left( \frac{d\varphi }{dt}\right) ^{2}$ , to
the energy scale, $U$ , of the spectrum of the rotational Hamiltonian, Eq.(%
\ref{Hrot}), we find for the angular velocity
\[
\left( \frac{d\varphi }{dt}\right) \sim U/\hbar =\frac{eH}{2M_{scol}c}
\]
with \
\begin{equation}
M_{scol}=\left( 2\left| \widetilde{g}\right| ^{2/3}\left| g\right| \right)
^{-1}m_{0}  \label{Mscol}
\end{equation}

This is an expression for an effective Larmor frequency for precession of
the entire spin texture about the external magnetic field axis, with an
effective mass, Eq.(\ref{Mscol}), which diverges with vanishing g-factor
like $g^{-5/3}$. \ For typical experimental values, $\widetilde{g}\sim .002$
, we find that\ $M_{scol}/m_{0}\sim 10^{4}$ .

In addition to the collective rotational motion of the entire spin texture
just described, the internal degrees of freedom of the spin texture can also
be excited, e.g. as spin waves associated with single electron-hole pair
excitations (spin-excitons) \cite{BIE81},\cite{KH84}. \ The above
consideration shows that for a sufficiently large Skyrmion the energy gap ,$%
\varepsilon _{sp}$, of the spin-waves is much larger than that of the
collective rotational spectrum. This separation of energy scales may be
expressed explicitly by writing the transverse electron spin density in the
form:
\begin{equation}
S_{+}\left( \mathbf{r},t\right) \equiv \frac{1}{4\pi }n\left( \mathbf{r}%
,t\right) =\frac{1}{4\pi }\widetilde{n}\left( \mathbf{r},t\right)
e^{i\varphi \left( t\right) }  \label{S+}
\end{equation}
where $\varphi \left( t\right) $ is the instantaneous collective rotation
angle, and $\widetilde{n}\left( \mathbf{r},t\right) $ stands for all the
other degrees of freedom in the electronic spin space. It can be derived by
expressing the phase of $n\left( \mathbf{r},t\right) =\left| n\left( \mathbf{%
r},t\right) \right| e^{i\theta \left( \mathbf{r},t\right) }$ \ as a Fourier
series $\theta \left( \mathbf{r},t\right) =\sum_{\mathbf{k}\neq \mathbf{0}%
}\theta _{\mathbf{k}}\left( t\right) e^{i\mathbf{k\cdot r}}+\theta _{\mathbf{%
0}}\left( t\right) $, and identifying the uniform term, $\theta _{\mathbf{0}%
}\left( t\right) $ , with $\varphi \left( t\right) $, so that $\widetilde{n}%
\left( \mathbf{r},t\right) =\left| n\left( \mathbf{r},t\right) \right| \exp %
\left[ \sum_{\mathbf{k}\neq \mathbf{0}}\theta _{\mathbf{k}}\left( t\right)
e^{i\mathbf{k\cdot r}}\right] $.

\section{Nuclear spin dynamics}

Let us, finally study in some detail the influence of these electron spin
excitations on the dynamics of nuclear spins via the hyperfine interaction
investigated in our model. The processes of nuclear spin relaxation and
decoherence are reflected in the time dependence of the average $%
I_{+,-}=\langle \widehat{I}_{+,-}\rangle $,where the brackets $\langle
...\rangle $ stand for the state of the combined system of the nuclear and
electronic spins ( see Ref.\cite{MBVW01}). Exploiting the adiabatic
approximation, which is valid when the effect of the hyperfine interaction
is so weak as to be neglected beyond the leading order, which is the first
order in the calculation of the nuclear spin eigen-energies, and the second
order in the calculation of relaxation and decoherence. Thus we have for the
transverse component of the nuclear spin located at $\mathbf{r}$, up to
second order of the corresponding perturbation theory \cite{MBVW01}:

\begin{eqnarray}
&&\hspace{2cm}\left[ \frac{\partial }{\partial t}+i\Omega \left( \mathbf{r}\right) \right]
I_{+}(\mathbf{r},t) =  \label{ddtI} \\
&&-\frac{\alpha ^{2}}{4}\int_{0}^{t}d\tau\left\langle 0\left| \left\{ \widehat{S}_{+}(\mathbf{r},t),\widehat{S}_{-}(%
\mathbf{r},\tau )\right\} \right| 0\right\rangle e^{i\varpi (\tau -t)}I_{+}(%
\mathbf{r},t)  \nonumber
\end{eqnarray}
where the symbol $\left\{ ,\right\} $ stands for anticommutator,
and the averaging is performed over the ground state $|0\rangle $
of the electronic system.The local NMR frequency $\Omega \left(
\mathbf{r}\right) $ corresponds to the unperturbed precession\ of
the nuclear spin in the
external static magnetic field (with the frequency $\varpi =\gamma _{n}B_{0}$%
) and the first order correction due to the local hyperfine interaction (
the Knight shift ), i.e. $\Omega \left( \mathbf{r}\right) =\gamma
_{n}B_{0}-\alpha \left\langle 0\left| \widehat{S}_{z}\left( \mathbf{r}%
\right) \right| 0\right\rangle $. Note that the corresponding correction due
to the transverse component of the hyperfine field is neglected in Eq.(\ref
{ddtI}). Note also that in the framework of the adiabatic approximation,
used in the derivation of Eq.(\ref{ddtI}), the weak time dependence of the
operator$\widehat{I}_{+}(\tau )$ ,due to depolarization, is neglected ( so
that $\widehat{I}_{+}(\tau )\simeq \widehat{I}_{+}(t)e^{i\varpi (\tau -t)}$
).

The resulting equation, (\ref{ddtI}) , is solved by
\begin{equation}
I_{+}(\mathbf{r},t)=I_{+}(\mathbf{r},0)e^{-\Gamma \left( \mathbf{r},t\right)
-i\Omega \left( \mathbf{r}\right) t}  \label{I+(t)}
\end{equation}
where
\[
\Gamma \left( \mathbf{r},t\right) =\text{Re}\int_{0}^{t}dt^{\prime }\xi
\left( \mathbf{r},t^{\prime }\right)
\]
and
\[
\xi \left( \mathbf{r},t\right) =\frac{\alpha ^{2}}{4}\int_{0}^{t}d\tau
e^{i\varpi (\tau -t)}\left\langle 0\left| \left\{ \widehat{S}_{+}(\mathbf{r}%
,t),\widehat{S}_{-}(\mathbf{r},\tau )\right\} \right| 0\right\rangle
\]

At filling factors slightly away from $\nu =1$ ,\ where the density of
Skyrmions is small and the interaction between them can be neglected, $%
\widehat{S}_{+}(\mathbf{r},t)$ may be written in the form (\ref{S+}),
describing a single Skyrmion centered at $\mathbf{r}=0$. \ On the large time
scale relevant to the nuclear spin dynamics of interest here,when the
internal degrees of freedom of the spin texture are essentially frozen, it
is possible to neglect the time dependence of $\widetilde{n}\left( \mathbf{r}%
,t\right) $ in Eq.(\ref{S+}) $\ $(by writing $\ \widetilde{n}\left( \mathbf{r%
},t\right) \approx \widetilde{n}\left( \mathbf{r}\right) $\ ) , so that:

\[
\xi \left( \mathbf{r},t\right) \approx \left( \frac{\alpha }{8\pi }\left|
n\left( \mathbf{r}\right) \right| \right) ^{2}\int_{0}^{t}d\tau e^{i\varpi
(\tau -t)}\left\langle 0\left| \left\{ e^{i\widehat{\varphi }\left( t\right)
},e^{-i\widehat{\varphi }\left( \tau \right) }\right\} \right|
0\right\rangle
\]
where $e^{i\widehat{\varphi }\left( t\right) }\equiv e^{it\widehat{H}%
_{rot}/\hbar }e^{i\varphi }e^{-it\widehat{H}_{rot}/\hbar }$. \ A
straightforward algebra yields:
\begin{equation}
e^{i\widehat{\varphi }\left( t\right) }=e^{i\varphi }\exp \left\{ i\frac{U}{%
2\hbar }t\left[ 1-2\left( i\frac{\partial }{\partial \varphi }+K\right) %
\right] \right\}  \label{Exphi(t)}
\end{equation}
so that the correlation function $\left\langle 0\left| \left\{ e^{i\widehat{%
\varphi }\left( t\right) },e^{-i\widehat{\varphi }\left( \tau \right)
}\right\} \right| 0\right\rangle =2\cos \left[ U\delta K\left( t-\tau
\right) /\hbar \right] $, where $\delta K\equiv \left[ K\right] -K$ , and $%
\left[ K\right] $ is the integer closest to $\left( K-1/2\right) $. \
Consequently, one finds that
\begin{equation}
\Gamma \left( \mathbf{r},t\right) =2\left( \frac{\alpha }{8\pi }\left|
n\left( \mathbf{r}\right) \right| \right) ^{2}\frac{1-\cos \left[ \left(
U\delta K/\hbar -\varpi \right) t\right] }{\left( U\delta K/\hbar -\varpi
\right) ^{2}}  \label{GamSk}
\end{equation}

This expression shows that as long as the rotational energy gap $U\left|
\delta K\right| $ is much larger than the nuclear Zeeman energy $\hbar
\varpi $, the off-diagonal element of the nuclear spin density matrix (i.e.
the coherence) does not decay, but oscillates very quickly (i.e. with
frequency $U\left| \delta K\right| /\hbar $ )\ \ between $I_{+}(\mathbf{r}%
,0) $ and $I_{+}(\mathbf{r},0)e^{-\left( A\left| S_{+}\left( \mathbf{r}%
\right) \right| /U\delta K\right) ^{2}}$ . \ It should be stressed that in
deriving Eq.(\ref{GamSk}) the interaction of the electronic system to its
environment was completely neglected. This coupling should lead to some
energy dissipation, which results in damping of the oscillatory component of
$\Gamma \left( \mathbf{r},t\right) $, so that for sufficiently long times, $%
I_{+}(\mathbf{r},t)\rightarrow I_{+}(\mathbf{r},0)e^{-\left( A\left|
S_{+}\left( \mathbf{r}\right) \right| /U\delta K\right) ^{2}}$.

As discussed above, the effective electron g-factor can become locally
sufficiently small to make the local Skyrmion radius large enough, so that
the corresponding rotational energy gap $U$ becomes comparable to the
nuclear Zeeman energy $\hbar \varpi $. \ For such a large Skyrmionic spin
texture the extremely slow collective spin rotation leads to a complete loss
of coherence of nuclear spins via the hyperfine coupling. Under this
condition the decay is Gaussian, $I_{+}(\mathbf{r},t)\sim $ \ $e^{-\left(
\alpha \left| n\left( \mathbf{r}\right) \right| /8\pi \right) ^{2}t^{2}}$,
with characteristic relaxation time $T_{2}\sim \hbar /A\left| S_{+}\left(
\mathbf{r}\right) \right| =\pi /K_{S}\left| \widetilde{S}_{+}\left( \mathbf{r%
}\right) \right| $, which is of the order of $0.1-1$ milliseconds for GaAs
MQW. It should be stressed here that the neglect of the first order
correction due to the transverse component of the hyperfine field in Eq.(\ref
{ddtI}) results in the vanishing of the equilibrium solution $I_{+}(\mathbf{r%
},t\rightarrow \infty )$. \ The present dynamical approach should be
therefore modified to take into account this correction in order to describe
relaxation to the nonvanishing nuclear spin texture, Eq.(\ref{I+(r)}).

\begin{figure}
\begin{center}
\includegraphics[width=7cm]{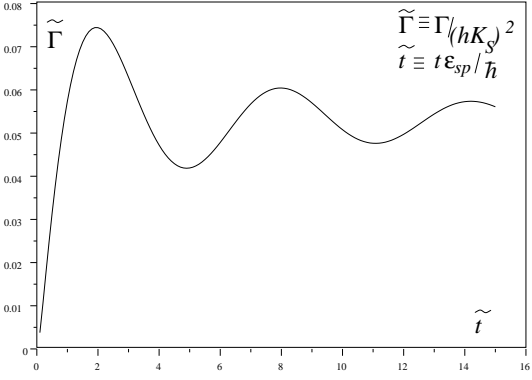}
\caption{ The real exponent of the coherence (off diagonal element of the
density matrix) of nuclear spin as a function of time in QH ferromagnet, \
for $\varepsilon _{C}/\varepsilon _{sp}=32$ (see text), showing incomplete
decoherence due to virtual electronic spin wave excitations.}

\end{center}
\end{figure}

At filling factor $\nu =1$, where the number of Skyrmions vanishes (note
that due to spatial inhomogeneity of the local filling factor some equal
number of Skyrmions and anti-Skyrmions can exist even at $\nu =1$ ), the
nuclear spin dynamics is controlled by the coupling to the well known gapped
spin waves. In the presence of the gap the virtual flip-flop excitations of
electronic spin waves via the hyperfine interaction ( which are the vacuum
quantum fluctuations of the QH ferromagnet) lead to decoherence of the
nuclear spin states, i.e.\cite{MBVW01}:\
\begin{eqnarray}
\Gamma \left( \mathbf{r},t\right) =\Gamma (t)=\left( hK_{S}\right)
^{2}\times  \nonumber \\
\int_{0}^{\infty }\widetilde{k}d\widetilde{k}e^{-\widetilde{k}^{2}/2}%
\frac{1-\cos ([\varepsilon _{ex}(\widetilde{k})/\hbar -\varpi ]t)}{%
[\varepsilon _{ex}(\widetilde{k})-\hbar \varpi ]^{2}}  \label{GamEx}
\end{eqnarray}
where $\varepsilon _{ex}(\widetilde{k})\approx \varepsilon _{sp}+\frac{1}{4}%
\varepsilon _{C}\widetilde{k}^{2}$ , for $\widetilde{k}=kl_{H}\ll 1$ , and $%
\varepsilon _{C}=\sqrt{\pi /2}\left( e^{2}/\kappa l_{H}\right) $ is the
Coulomb energy. \ Similar to the case of the collective mode with the large
excitations gap, discussed below Eq.(\ref{GamSk}), in the present case the
coherence does not decay to zero at any time. In contrast to the effect of
the undamped collective mode, however, the presence of a continuous band of
spin waves above the Zeeman gap $\varepsilon _{sp}$ results in some
irreversible loss of coherence. This decoherence occurs on a very short time
scale:- the precession period of the electronic spin, $2\pi /\omega _{sp}$,
whereas for longer times the coherence undergoes damped oscillation\ (with
frequency $\omega _{sp}$ ) about a nonzero value \cite{Palma96} (see Fig.(1)
), that is: $I_{+}(\mathbf{r},t)e^{i\Omega \left( \mathbf{r}\right)
t}\rightarrow I_{+}(\mathbf{r},0)\exp \left[ -2\left( \varepsilon
_{C}/\varepsilon _{sp}\right) \left( hK_{S}/\varepsilon _{sp}\right) ^{2}%
\right] $. \ For $hK_{S}\ll \varepsilon _{sp}\left( \varepsilon
_{C}/\varepsilon _{sp}\right) ^{1/2}$ (e.g. for GaAs MQW $hK_{S}/\varepsilon
_{sp}\sim 10^{-7}$ , and $\varepsilon _{C}/\varepsilon _{sp}\sim 30$ at $%
H=10 $ T ), the corresponding decoherence is negligibly small. \

In actual heterojunctions the electronic Zeeman gap is usually much smaller
than the theoretical value. It can be further suppressed by applying
pressure \cite{Maude96}, so that the situation of gapless spin waves may not
be unrealistic experimentally. In this case the integral over $\widetilde{k}$
in Eq.(\ref{GamEx}) becomes in the long time limit $t\gg \hbar /\varepsilon
_{C}$ :
\begin{eqnarray}
\Gamma (t)&=&2\left( hK_{S}\right) ^{2}\int_{0}^{\infty }\widetilde{k}d%
\widetilde{k}e^{-\widetilde{k}^{2}/2}\frac{\sin ^{2}(\varepsilon _{C}%
\widetilde{k}^{2}t/8\hbar )}{(\varepsilon _{C}\widetilde{k}^{2}/4)^{2}} \nonumber \\
&\rightarrow& \left( \frac{\left( 2\pi \right) ^{2}K_{S}^{2}}{\varepsilon
_{C}/h}\right) t \nonumber
\end{eqnarray}
so that the decay of coherence with time is a simple exponential, $I_{+}(%
\mathbf{r},t)e^{i\Omega \left( \mathbf{r}\right) t}\rightarrow I_{+}(\mathbf{%
r},0)\exp \left( -t/T_{2}\right) $ , \ where
\[
T_{2}=\left[ \frac{\left( \varepsilon _{C}/\varepsilon _{sp}\right) }{\left(
2\pi \right) ^{2}\left( hK_{S}/\varepsilon _{sp}\right) ^{2}}\right] \left(
\frac{2\pi }{\omega _{sp}}\right)
\]
For GaAs MQW this expression yields $T_{2}\sim \allowbreak 10^{3}$ sec. ,
indicating that the long relaxation times observed experimentally in the QH
ferromagnetic state can be reasonably explained by a gapless spin exciton
spectrum.

\section{Conclusion}

In this paper it was demonstrated how a coherent superposition of many
nuclear spin states can be prepared and manipulated via the hyperfine
interaction by varying the LL filling factor in appropriately designed QH
Ferromagnet. During the manipulation periods the electronic spins form
spatially large spin textures, where the average spin polarization in the
plane perpendicular to the external magnetic field varies smoothly, and the
individual spins are strongly correlated over large microscopic regions. The
nuclear spins, which are coupled to their environment only via the hyperfine
interaction with the electron spins, follow the changes in the electronic
spins system by creating their own spin textures,which replicate the
electronic ones. This effect is expected to be significant only in very
special systems, where the strength of the hyperfine interaction is
comparable to the nuclear Zeeman energy. \ The nuclear spins relaxation and
decoherence processes in such states are governed by the coupling to
collective spin rotational modes of the entire electronic spin textures,
which have vanishingly small excitation gap in regions where the local
electronic g-factor vanishes.

\ It turns out that GaAs MQW, despite its remarkable features described
above, is not suitable for our purpose. The reason is twofold:

1) The hyperfine coupling constant in GaAs is much too small to be effective
in manipulating nuclear spins in the QW.

2) The nuclear spin dephasing time in quantum well structures based on
GaAS/AlGaAs, is expected to be much smaller than the shortest value of $%
T_{1} $ found in this experiments. This drawback is due to the fact that all
abundant isotopes in this compound (i.e. $^{69}$Ga , $^{71}$Ga , $^{75}$As ,
all with $I=3/2$ , and $^{27}$Al with $I=5/2$ ) have non-zero nuclear spins,
so that significant dephasing due to dipolar interactions is expected.
Indeed, a rough estimate for $T_{2}$ for a solid in which each nuclear spin
has nearby nuclear spins is in the range of milliseconds \cite
{Slichter90,Mieher66}

A possible solution for both problems may be found in MQW structures
composed of Si/Si$_{1-x}$Ge$_{x}$ \cite{shlimak83}: The most abundant
isotopes of these nuclei have zero nuclear spins , so that by purifying the
host sample isotopically \cite{shlimak83}, and then weakly doping with, e.g.
$^{31}$P donor \cite{Kane98}, which has $I=1/2$, one may reduce the dipolar
dephasing to the desired low level.

Furthermore, the hyperfine coupling between the conduction electrons and the
$^{31}$P nucleus in the Si host is strongly enhanced, due to the high
concentration of the electron s-orbitals at the donor nucleus. Thus, a
Knight shift of about $30$ $\text{MHz}$, which is comparable to the NMR
frequency at about $1\text{T}$, can be obtained for Si:$^{31}$P \cite{Kane98}%
.

\textbf{Acknowledgments}: We thank Sean Barret and D. Maude for illuminating
discussions concerning the experimental aspects of this paper. This research
was supported by the fund for the promotion of research at the Technion.


\begin{thebibliography}{99}
\bibitem{Steane98}  For a review see: A. Steane, Rep. Prog. Phys., \textbf{61%
}, 117 (1998).

\bibitem{PVK98}  V. Privman, I.D.\ Vagner and G.\ Kventsel, Phys. Let. A
\textbf{236}, 141 (1998).

\bibitem{Barret95}  S.E.Barret, G.Dabbagh, L.N.Pfiffer, K.W.West , and
R.Tycko, Phys. Rev. Lett. \textbf{74} , 5112 (1995)

\bibitem{Tycko95}  R. Tycko, S. E. Barrett, G. Dabbagh, L. N. Pfeiffer, and
K. W. West, Science \textbf{268 }, 1460 (1995).

\bibitem{Girvin99}  For a review see S. M. Girvin, cond-mat/9907002 (1999).

\bibitem{MBVW01}  T. Maniv, Yu. A Bychkov, I.D. Vagner, P. Wyder, Phys. Rev.
B \textbf{64}, 193306 (2001).

\bibitem{Sondhi93}  S. L. Sondhi, A. Karlhede, S. A. Kivelson, and E. H.
Rezayi, Phys. Rev. B \textbf{47}, 16419 (1993).

\bibitem{BIE81}  Yu. A. Bychkov, S. V. Iordanskii and G. M. Eliashberg,
JETP\ Lett. \textbf{33}, 143 (1981)

\bibitem{KH84}  C. Kallin and B. I. Halperin, Phys. Rev. B \textbf{31} ,
3635 (1984).

\bibitem{BMVW94}  Yu. A. Bychkov, T. Maniv. I.D. Vagner and P. Wyder, Phys.
Rev. Lett. \textbf{73}, 2911 (1994); T. Maniv, Yu. A. Bychkov, A.
Kaplunovsky and I.D. Vagner, Physica B \textbf{204}, 134 (1995); Yu. A.
Bychkov, T. Maniv. I.D. Vagner and P. Wyder, Europhys. Lett. \textbf{40},
557 (1997).

\bibitem{Fertig94}  H. A. Fertig, L. Brey, R. Cote, and A. H. MacDonald,
Phys. Rev. B \textbf{50} , 11018 (1994);L. Brey, H. A. Fertig, R. Cote, and
A. H. MacDonald, Phys. Rev. Lett. \textbf{75}, 2562 (1995).

\bibitem{BMV96}  Yu. A. Bychkov, T. Maniv and I.D. Vagner, Phys. Rev. B
\textbf{53}, 10148 (1996).

\bibitem{BKMV98}  Yu. A. Bychkov, A. V. Kolesnikov, T. Maniv and I.D.
Vagner, J. Phys.: Cond. Matt. \textbf{10}, 2029 (1998).

\bibitem{MFB96}  A. H. MacDonald, H. A. Fertig, and L. Brey, Phys. Rev.
Lett. \textbf{76} , 2153 (1996).

\bibitem{FBCMKS97}  H. A. Fertig, L. Brey, R. Cote, and A. H. MacDonald, A.
Karlhede , and S. I. Sonhi, Phys. Rev. B \textbf{55}, 10671 (1997).

\bibitem{CMBFGS97}  R. Cote, A. H. MacDonald, L. Brey, H. A. Fertig, S. M.
Girvin, and H. T. C. Stoof, Phys. Rev. Lett., \textbf{78}, 4825 (1997); A.
G. Green, I. I. Kogan, and A. M. Tsvelik, Phys. Rev. B \textbf{54}, 16838
(1996).

\bibitem{Barrett01}  P. Khandelwal, A.E. Dementyev, N.N. Kuzma, S.E.
Barrett, L.N. Pfeiffer and K.W. West, Phys. Rev. Lett. \textbf{86}, 5353
(2001).

\bibitem{Nederveen99}  A.J. Nederveen and Yu.V. Nazarov, Phys. Rev. Lett.
\textbf{82}, 406 (1999).

\bibitem{Unruh95}  W. G. Unruh, Phys. Rev. B \textbf{51} , 992 (1995)

\bibitem{Palma96}  G. M. Palma, K. A. Suominen and A. K. Ekert, Proc. Royal
Soc. London, A \textbf{452 }, 567 (1996).

\bibitem{Maude96}  D.K. Maude, M. Potemskii, J.C. Portal, M. Henini, L.\
Eavas, G. Hill, and M.A. Pate, Phys. Rev. Lett. \textbf{77}, 4604 (1996).

\bibitem{Slichter90}  C. P. Slichter, \textit{Principles of Magnetic
Resonance }, 3rd Ed. (Springer-Verlag, Berlin, 1990).

\bibitem{Mieher66}  R.L. Mieher, in \textit{Semiconductors and Semimetals }%
,ed. R.K. Willardson and A.C.Beer, \textbf{2}, Acad. Press (1966).

\bibitem{shlimak83}  I. S. Shlimak \textit{et al}., Sov. Techn. Phys. Lett.,
\textbf{9}, 377 (1983)

\bibitem{Kane98}  B. E. Kane, Nature, \textbf{393 },133 (1998)

\end{thebibliography}
\end{document}